\documentclass{emulateapj}
\begin{document}

\title{A New Method for Obtaining the Star Formation Law in Galaxies}

\author{Jonathan S. Heiner\altaffilmark{1}, Ronald J. Allen\altaffilmark{2}, \& Pieter C. van der Kruit\altaffilmark{3}
}
\altaffiltext{1}{D\'{e}partement de physique, de g\'{e}nie physique et d'optique, Universit\'{e} Laval, Qu\'{e}bec, QC G1V 0A6, Canada}
\altaffiltext{2}{Space Telescope Science Institute, Baltimore, MD 21218, USA}
\altaffiltext{3}{Kapteyn Astronomical Institute, University of Groningen, PO Box 800, 9700 AV Groningen, the Netherlands}
\email{jonathan.heiner.1@ulaval.ca}

\begin{abstract}
We present a new observational method to evaluate the exponent of the star formation law as initially formulated by Schmidt, i.e.\ the power-law expression assumed to relate the rate of star formation in a volume of space to the local total gas volume density present there. 

Total volume densities in the gas clouds surrounding an OB association are determined with a simple model which considers the atomic hydrogen as a photodissociation product on the cloud surfaces. The photodissociating photon flux incident on the cloud is computed from the far-UV luminosity of the OB association and the geometry. As an example, we have applied this ``PDR Method'' to a sample of star-forming regions in M33 using \textit{VLA} 21-cm data for the \mbox{H{\sc i}} and \textit{GALEX} imagery in the far-UV. With these two observables, our approach provides an estimate of the total volume density of hydrogen (atomic + molecular) in the gas clouds surrounding the young star cluster. 

A graph in logarithmic coordinates of the cluster UV luminosity \textit{versus} the total density in the surrounding gas provides a direct measure of the exponent of the star formation law. However, we show that this plot is severely affected by observational selection, which renders large areas of the diagram inaccessible to the data. An ordinary least-squares regression fit to a straight line therefore gives a strongly biased result. In the present case, the slope of such a fit primarily reflects the boundary defined when the 21-cm line becomes optically thick and is no longer a reliable measure of the \mbox{H{\sc i}} column density. We use a  maximum-likelihood statistical approach which can deal with truncated and skewed data, and also takes account of the large uncertainties in the total gas densities which we derive. The exponent we obtain for the Schmidt law in M33 is 1.4 $\pm$ 0.2.
\end{abstract}

\keywords{galaxies: individual (M33) --- ultraviolet: galaxies --- galaxies: ISM --- ISM: clouds --- ISM: molecules --- ISM: atoms}

\maketitle

\section{Introduction}

Some 50 years ago, Maarten Schmidt published a pair of seminal papers titled ``The Rate of Star Formation'' that discussed the relationship between the interstellar gas and the present and past rates of star formation in the Galaxy \citep[][]{1959ApJ...129..243S,1963ApJ...137..758S}. This was a topic of considerable interest at the time, since radio telescopes had begun to map the distribution of atomic hydrogen over the Galaxy in some detail. Schmidt postulated a simple power-law relationship between the volume density of the interstellar gas in a region of the Galaxy and the number of stars per unit volume and time formed there. He wrote this postulate as:
\begin{equation}
  \mbox{SFR} \propto n^\alpha,
  \label{eqn:sfr}
\end{equation}
where SFR is the local star formation rate e.g.\ in solar masses per cubic parsec per year, and $n$ is the total volume density of the interstellar gas. For the former, Schmidt used the Z-distribution and counts of young stars. For the latter, he used estimates of the density and Z-distribution of atomic hydrogen, while at the same time acknowledging that these values might be wrong owing to the unknown amount and Z-distribution of molecular hydrogen which may also be present. With these simplifying assumptions, Schmidt found that the exponent $\alpha$ in equation \ref{eqn:sfr} was approximately 2 in the local Galactic neighborhood. 

The difficulty of obtaining estimates of total gas volume density in the interstellar medium (ISM) has prevented further attempts to obtain estimates of the star formation law as initially formulated by Schmidt. Instead, the suggestion was made to relate a star formation rate \textit{per surface area} to the \textit{surface density} of atomic hydrogen as determined from the surface brightness in the 21-cm \mbox{H{\sc i}} line \citep[e.g.][]{1989A&A...223...42B}. While this approach necessarily ignores differences in the line-of-sight distributions of the gas and young stars, it is straighforward to implement. For the last $\approx 20$ years this approach has been the mainstay of observational studies of the star formation law in disk galaxies at low inclinations, and the formulation is now called the ``Kennicutt-Schmidt Law'' in recognition of the extensive work in this area by R. Kennicutt. These studies are summarized in \citet{1998ARA&A..36..189K}, where it is concluded that the exponent $a$ equals 1.4 in an expression for surface densities of the form $\Sigma_{SFR} \propto \Sigma_{gas}^a$. It is then generally assumed that the $a$ determined from surface densities is the same as the $\alpha$ for volume densities in our equation \ref{eqn:sfr}. In more recent studies the 21-cm \mbox{H{\sc i}} data has been augmented with estimates of the molecular content using the surface brightness of the CO(1-0) line as a tracer, but the general approach has otherwise remained the same. \citet{2009MNRAS.396.1579D} have recently summarized the subject and shown that a simple physical model of star formation in the ISM is capable of reproducing a power-law form for the Kennicutt-Schmidt law.

The existence of a power-law form for the star formation has been confirmed in many galaxies, although a wide range of values has been reported \citep{1997ASSL..161..171K}. The recent addition of CO as a tracer for the molecular gas has not reduced the scatter in the results. For instance in M33, \citet{2004ApJ...602..723H} find a slope of 1.38 for the molecular gas (based on CO measurements) but 3.3 for the total gas relation (with a monotonically-increasing atomic gas fraction at increasing galactocentric distances) and \citet{2010A&A...510A..64V} report values varying from 1.1 to 2.9. A number of concerns can be expressed about the meaning of these results given the input data. For instance, what are the consequences of averaging along the line of sight when we know from Galactic studies that the spatial distributions of the tracers are not the same on scales below a few hundred parsec? And how reliable are the tracers we are using for the gas (the beam-smoothed 21-cm and CO(1-0) line surface brightnesses) in providing column densities independent of local physical conditions and radiative transfer effects?

In view of such concerns and of the wide range of values reported for the exponent in the Kennicutt-Schmidt Law, it seems useful to develop other methods for examining the quantitative relationship between gas and young stars in galaxies. In this paper, we present a new method based on treating the \mbox{H{\sc i}} found near OB associations as arising in photo-dissociation regions (PDRs) which develop on the surfaces of the parent giant molecular clouds (GMCs) under the action of the far-UV radiation produced by nearby young stars. This approach, which we call the ``PDR method'',  provides us with a way of estimating the \textit{total} gas volume density (atomic + molecular) in the parent GMCs. The far-UV luminosity of the nearby OB association is a measure of the star formation rate at that location in the galaxy, so that a diagram of these luminosities vs.\ the estimates of volume densities in the surrounding clouds (of which there may be several) is amenable to interpretation as a star formation law in the sense originally described by Schmidt.

We describe the PDR method in more detail in \S \ref{sec:pdrmethod}. The data used, and our approaches to dealing with the effects of observational selection, are detailed in \S \ref{sec:data}. We estimate the exponent of the star formation law in M33 in two ways in \S \ref{sec:results}. Our conclusions are summarized in \S \ref{sec:conclusions}.

\section{The PDR method}
\label{sec:pdrmethod}

Molecular hydrogen is difficult to observe directly in the ISM, owing primarily to its lack of a dipole moment. The most common indirect means of inferring its presence in the ISM is to observe the low-level rotational lines of the carbon monoxide (CO) molecule. The alternative method we present here is based on the physics of photodissociation of molecular hydrogen and the discovery by \citet{1986Natur.319..296A} that this process was responsible for the production of atomic hydrogen in the prominent spiral arms of M83. The method was first featured in \citet{1997ApJ...487..171A}, then in \citet{2000ApJ...538..608S}, \citet{2008ApJ...673..798H}, and \citet{2008A&A...489..533H}. The method provides estimates of the \textit{total} hydrogen volume densities (atomic + molecular) in gas clouds located in close proximity to regions of recent star formation. These clouds may be considered representative of the parent GMCs out of which the young stellar association formed. In \citet{2009ApJ...700..545H} we presented the first results of applying the PDR method to a study of GMCs in M33 at a linear resolution of 20 pc, the highest available in our observations. Here we use the data from a larger survey of clouds (Heiner et al., in preparation) at a linear resolution of $\approx 80$ pc to specifically address the star formation law in this galaxy. We begin by describing the geometry and reviewing the relevant physics of our model.

\subsection{A simple PDR model}
\label{sec:model}

\begin{figure*}[ht]
  \centering
  \includegraphics[width=0.8\linewidth]{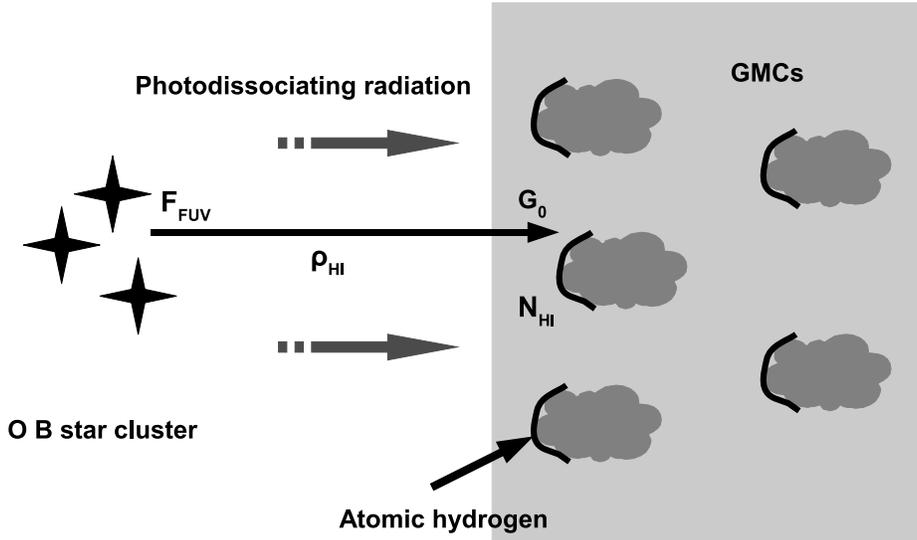}
  \caption{\label{fig:pdrschematic} A schematic view of a typical region with candidate PDRs. The OB star cluster floods the surrounding highly-porous ISM with FUV radiation, and layers of atomic hydrogen are formed on the surfaces of the remaining molecular clouds.}
\end{figure*}

Clusters of young OB stars are copious producers of far-UV (FUV) photons in galaxies, and the FUV luminosity of the stellar association is a measure of the local star formation rate \citep[see e.g.\ ][]{1998ARA&A..36..189K}. The FUV photons also radiate into the surrounding interstellar medium and dissociate some of the molecular gas found in the remaining GMCs. Owing to the high porosity of the ISM, the dissociating photons penetrate hundreds of parsecs into the interstellar medium \citep[as confirmed by e.g.\ ][]{2009ApJ...700..545H}, creating ``skins'' of atomic hydrogen on the surfaces of the GMCs which they encounter; see Figure \ref{fig:pdrschematic}. As we shall show, the \mbox{H{\sc i}} column densities so produced on the surfaces of the GMCs can be directly related to the total hydrogen volume density in the cloud; that gas will be mostly atomic on the cloud surfaces, but predominantly molecular deep inside the GMCs. The resulting total gas volume densities obtained for samples of GMCs surrounding a selection of OB associations is then combined together with the local FUV-determined star formation rate for each association to obtain an estimate for the local star formation law in the galaxy.

A key step in the modeling is to use the physics of \mbox{H$_2$} dissociation by FUV photons in order to relate the observed \mbox{H{\sc i}} column densities on the cloud surfaces to the total gas densities in the cloud. We use a simple ``slab'' model described initially by \citet{1988ApJ...332..400S} and in more detail by  \citet{2004ASSL..319..731A}, with improved coefficients provided by \citet{2004ApJ...608..314A}.

The \mbox{H{\sc i}} column density in a PDR is calculated with the same physics used to determine the excitation of the \mbox{H$_2$} near-infrared fluorescence lines \citep{1988ApJ...332..400S}. We use the formulation in Appendix A of \citet{2004ApJ...608..314A}. Briefly, the model is a simple semi-infinite slab geometry in statistical equilibrium with FUV radiation incident on one side, and a \mbox{H{\sc i}} $\leftrightarrow$  \mbox{H$_2$} dissociation-reformation equilibrium in the slab on the right side.  The solution appropriate for our present purposes gives the steady state \mbox{H{\sc i}}\  column density along a line of sight perpendicular to the face of the slab as a function of $G_0$, the incident UV intensity scaling factor \citep[see Appendix B of][for a definition of $G_0$]{2004ApJ...608..314A}, and the total volume density $n$
of H nuclei in the slab. The result is:
\begin{equation}
N(\mbox{H{\sc i}})={1 \over \sigma} \times \ln{\left[ {{D S} \over {R n}} G_0
+ 1 \right]},
\label{eqn:nhi}
\end{equation}
where $D$ is the unattenuated \mbox{H$_2$} photodissociation rate in the average interstellar radiation field, $R$ is the \mbox{H$_2$} formation rate coefficient on grain surfaces, $\sigma$ is the effective grain absorption cross section per H nucleus in the FUV continuum, $G_0$ is the incident UV intensity scaling factor, $N(\mbox{H{\sc i}})$ is the \mbox{H{\sc i}} column density, and $n$ is the volume density of H nuclei in the slab. This equation has been developed using a simplified three-level model \citep{1988ApJ...332..400S} for the excitation of the \mbox{H$_2$} molecule. It is applicable for low-density ($n \lesssim 10^4$ cm$^{-3}$), cold (T $\lesssim 500$ K), isothermal, and static conditions, and neglects contributions to $N(\mbox{H{\sc i}})$ from ion chemistry and direct dissociation by cosmic rays.  The quantity $S$ is a dimensionless function of the effective grain absorption cross section $\sigma$, the absorption self--shielding function $f$, and the column density of molecular hydrogen $N_2$:
\begin{displaymath}
S = \int_0^{N_2} \sigma f e^{-2\sigma N_2^\prime} {\rm d}N_2^\prime.
\end{displaymath}
The function $S$ becomes constant for large values of $N_2$ due to self--shielding \citep{1988ApJ...332..400S}.  Using the parameter values in this equation adopted by \citet{2004ApJ...608..314A}, and writing explicitly the dependence on the dust-to-gas ratio \citep{2004ASSL..319..731A}, we have:
\begin{equation}
N(\mbox{H{\sc i}}) = \frac{7.8 \times 10^{20}}{\delta/\delta_0} \ln\left[1+\frac{106 \times G_0}{n
\sqrt{\delta/\delta_0}} \right],
\label{eqn:n}
\end{equation} 
where $N(\mbox{H{\sc i}})$ is $N_1$, the (background subtracted) atomic hydrogen column density (in $\rm{cm^{-2}}$), $\delta/\delta_0$ is the dust-to-gas ratio in the ISM expressed in terms of the value $\delta_0$ in the neighborhood of the sun, $G_0$\ is the (background subtracted) incident FUV flux calculated at the affected \mbox{H{\sc i}} patch, and $n = n_1 + 2 n_2$ is the total baryonic gas volume density (atomic + molecular). This gas will be mostly atomic on the surface of the GMC and mostly molecular deep inside the cloud. If we can measure $N(\mbox{H{\sc i}})$, $\delta/\delta_0$, and $G_0$, equation \ref{eqn:n} can be inverted to solve for $n$; this is the essence of our approach to obtaining volume densities in interstellar clouds. Although the exponential form of the result provides only a noisy estimate of the volume density, many measurements of $n$ can be made on GMCs in the immediate neighborhood of an OB association, thereby improving the overall precision.

Note that the method we have described does not make use of the observed \mbox{H{\sc i}} column density  to calculate the gas volume density from some estimated (but not measured) thickness. Such an estimate of \mbox{H{\sc i}} \textit{volume} density $n_1$ may, however, provide a useful ``lower limit'' check on the value of  $n = n_1 + 2 n_2$ calculated by the PDR method, since the transition region from \mbox{H{\sc i}} to \mbox{H$_2$} is known to be very abrupt in most PDRs.

To summarize, the steps involved in applying the PDR method are:
\begin{itemize}
\item Identify the sample of OB star clusters in the galaxy around which candidate PDRs can be sought. The clusters are knots of bright UV emission, and the candidate PDRs are patches of \mbox{H{\sc i}} surrounding them.
\item Determine the FUV flux of the OB clusters, e.g.\ on \textit{GALEX} FUV images. The effective wavelength of the \textit{GALEX} FUV band is $\approx 150$ nm, and since the spectrum is expected to be quite flat in this wavelength range the observed \textit{GALEX} FUV is a reasonable proxy for dissociating radiation at $\approx 100$ nm \citep[see][]{1978ApJS...36..595D}.
\item Identify the associated \mbox{H{\sc i}} patches on the surfaces of the surrounding GMCs, measure their surface brightnesses on \textit{VLA} 21-cm images, and convert those brightnesses to \mbox{H{\sc i}} column densities. Here we need to be cognizant of the fact that, with our relatively high spatial resolution, we may finally be resolving the \mbox{H{\sc i}} features, thus becoming sensitive to optical depth effects. Indeed, evidence for the presence of optically-thick \mbox{H{\sc i}} features in our M33 data will be presented later. 
\item Measure, and as far as possible de-project, the separation between the central UV source(s) and the \mbox{H{\sc i}} patches. This requires that the \textit{GALEX} and \textit{VLA}-\mbox{H{\sc i}} images be accurately aligned.
\item Obtain (or estimate) dust-to-gas ratios in the gas, ideally local to the candidate PDR.
\end{itemize}
This procedure allows us then to calculate the total hydrogen volume densities as local spot measurements at the location of candidate PDRs.

\section{Data}
\label{sec:data}

The \mbox{H{\sc i}} observations of M33 used in this paper were provided by David Thilker and Rob Braun (2007, private communication) and are presented and discussed in Heiner et al. (in preparation). We used the PDR method on these data to calculate the total hydrogen volume densities used here.
The linear resolution of the \mbox{H{\sc i}} data is $\approx 80$ pc. The local star formation rate at the location of the candidate PDRs is estimated using the far-UV luminosity of the parent OB associations from the \textit{GALEX} data, assuming that the star formation rate is directly proportional to the luminosity at the \textit{GALEX} far-UV wavelength \citep[see][]{1998ARA&A..36..189K,1998ApJ...498..106M}). In that case, using the far-UV luminosity results in the same power law slope when relating $L_{UV}$ to $n$. In the absence of region-specific extinction measurements, we have not made any corrections for extinction; a global extinction would merely shift all log-luminosities equally, and therefore would not influence the calculated value of the exponent in the star formation law. The volume Schmidt Law correlation we seek is of the form:
\begin{equation}
  \log{L_{UV}} \propto \alpha \log{n}.
  \label{eqn:logln}
\end{equation}
%

%Figure positioned manually
\begin{figure*}[t]
  \centering
  \includegraphics[width=\linewidth]{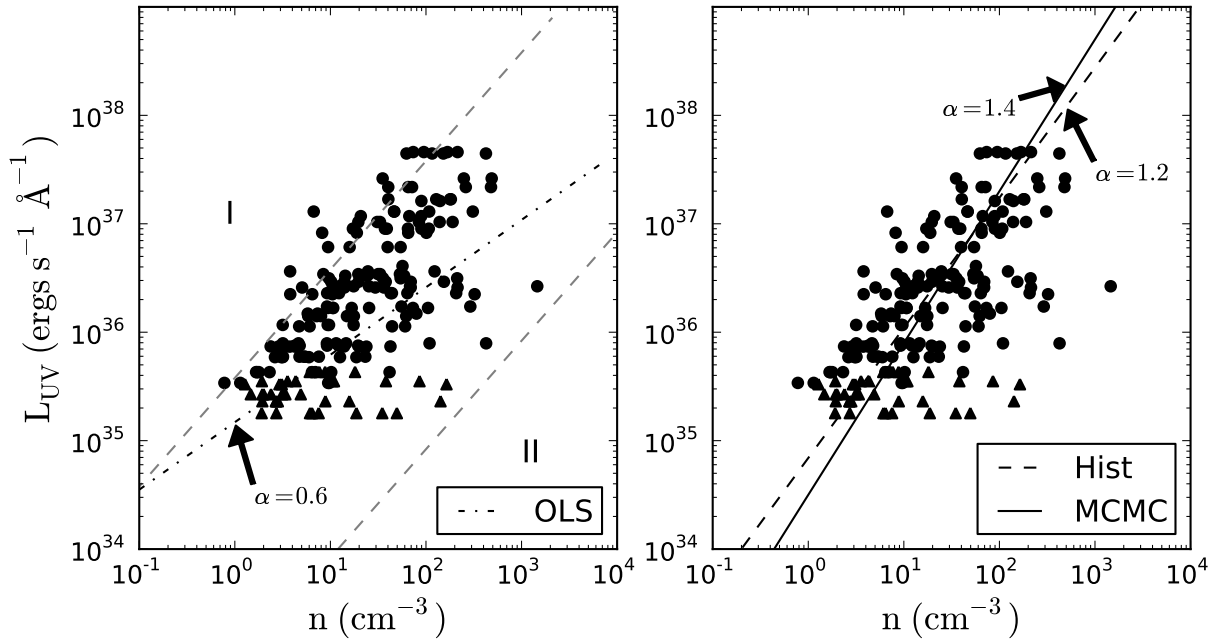}
  \caption{\label{fig:m33sfr} The far-UV luminosity of our candidate PDRs is plotted against the total hydrogen volume density. \textbf{Left panel:} The selection effects are indicated with dashed lines bounding areas where the \mbox{H{\sc i}} column density upper limit (I) and the \mbox{H{\sc i}} column density lower limit (II) are found. The ordinary least-squares (OLS) fit is represented by the dashed-dotted line, which is $L_{UV} \propto n^{0.6}$. \textbf{Right panel:} The dashed line labeled ``Hist'' graphs $L_{UV} \propto n^{1.2}$, obtained using histograms of the distribution of $n$; see text for further explanation. The solid line labeled ``MCMC'' corresponds to $L_{UV} \propto n^{1.4}$, from the maximum likelihood estimate. This last value of 1.4 (with an estimated uncertainty of 0.2) is our best estimate of the exponent in the Schmidt Law for star formation in M33.}
\end{figure*}

\section{Results}
\label{sec:results}

The computed values of $L_{UV}$ and $n$ from  Heiner et al. (in preparation) are shown in Figure \ref{fig:m33sfr} in log coordinates, along with several fits to straight lines. This figure shows a rough correlation, as might be expected from a star formation law of the form of equation \ref{eqn:logln}. The circles correspond to regions inside $R_{25}$, whereas the triangles are candidate PDRs outside of $R_{25}$. We have made no attempt to fit these regions separately. The straightforward ordinary least-squares (OLS) fit to the data represented by the dash-dotted line in the left panel yields $\alpha = 0.6$.

\subsection{Selection effects}
\label{sec:selection}

The data plotted in Figure \ref{fig:m33sfr} clearly avoid large regions of the graph, as follows:

\subsubsection{Low values of $L_{UV}$}

Our survey of star clusters in M33 cuts off below a limit of $L_{UV} \approx 2 \times 10^{35}$ ergs/sec/\AA). Most of these faint clusters (which contain few O stars) are found outside the $R_{25}$ isophotal radius in regions of low confusion; here the lower limit is set by the sensitivity of the \textit{GALEX} data. Such faint clusters also exist inside $R_{25}$ (along with a profusion of widely-scattered FUV-producing B stars), but confusion prevents their accurate tally. However, complete counts are not required for our study, since we subtract away the mean background levels on the FUV and \mbox{H{\sc i}} images in order to establish the excess $L_{UV}$ of the cluster and the excess \mbox{H{\sc i}} which the cluster's FUV photons produce on the surfaces of nearby GMCs. In this paper we have therefore not pursued FUV-faint objects inside the main disk of the galaxy\footnote{A quantitative study of these ``backgrounds'' is nevertheless highly relevant to the interesting question of what \textit{fraction} of the total \mbox{H{\sc i}} content of the entire galaxy arises through photodissociation, i.e., what fraction of the \mbox{H{\sc i}} present is not ``primordial'', but rather a \textit{product} of the star formation process.}.

\subsubsection{High values of $L_{UV}$}

At the bright end of the FUV luminosity we see that there are no OB associations more luminous than NGC 604 at $L_{UV} \approx 5 \times 10^{37}$, the equivalent of several dozens of O4 stars \citep[as inferred from][]{2003AJ....125.3082B}. This limit is apparently set by the physics of ``remaining'' steps in the star formation process in M33, which we do not consider here (nor even pretend to understand).

\subsubsection{Region II}

The sloping dashed line (partially) bounding the empty triangular region labelled II in the lower right corner of the left panel of Figure \ref{fig:m33sfr} indicates an \mbox{H{\sc i}} column density lower limit of $\approx 6 \times 10^{19}\ \rm{cm}^{-2}$, at a typical dust-to-gas ratio of around 0.33. The available data progressively disappears as we approach this line, which marks the approximate sensitivity limit of our \textit{VLA}-\mbox{H{\sc i}} data. This is a common (but largely benign) observational selection effect; the computed values of $n$ are more noisy here, but they are not strongly biased. A small bias could creep in if the slope of the line reflecting the selection effect were significantly different from the slope of the actual star formation relation, and at the same time there is an abundance of data points near this selection limit. Although the former is likely to be true, the latter is most assuredly not. 

\subsubsection{Region I}

The other empty triangular region labeled I in the upper left corner of the left panel of Figure \ref{fig:m33sfr} appears to arise from a more interesting (and also a more sinister) observational selection effect. The sloping dashed boundary line here is defined by an \mbox{H{\sc i}} column density upper limit of $5 \times 10^{21}\ \rm{cm}^{-2}$ at a characteristic dust-to-gas ratio; higher column density estimates are apparently very rare. We suggest that this is a consequence of the 21-cm line becoming progressively more optically thick at about this limiting value, thereby underestimating the true \mbox{H{\sc i}} column density and hence, following equation \ref{eqn:n}, \textit{over-}estimating the volume densities $n$. Points can lie somewhat to the left of this boundary owing, for example, to different dust-to-gas ratios compared to the typical value of 0.25 that we adopted. The different typical dust-to-gas ratio from the one we used to delineate region II (0.33) is a reflection of the fact that the higher column densities occur in the inner regions of M33, where the metallicity is higher as well. On the other hand, the lower column densities are measured mostly in the outer regions, where the metallicity is systematically lower. This selection effect is an optical depth limit, appearing as a consequence of the improved spatial resolution of the present data set that leads to a (partial) resolution of the M33 GMCs in \mbox{H{\sc i}}. 

\subsection{A simple correction}
\label{sec:correction}

\begin{figure*}[ht!]
  \centering
  \includegraphics[width=\linewidth]{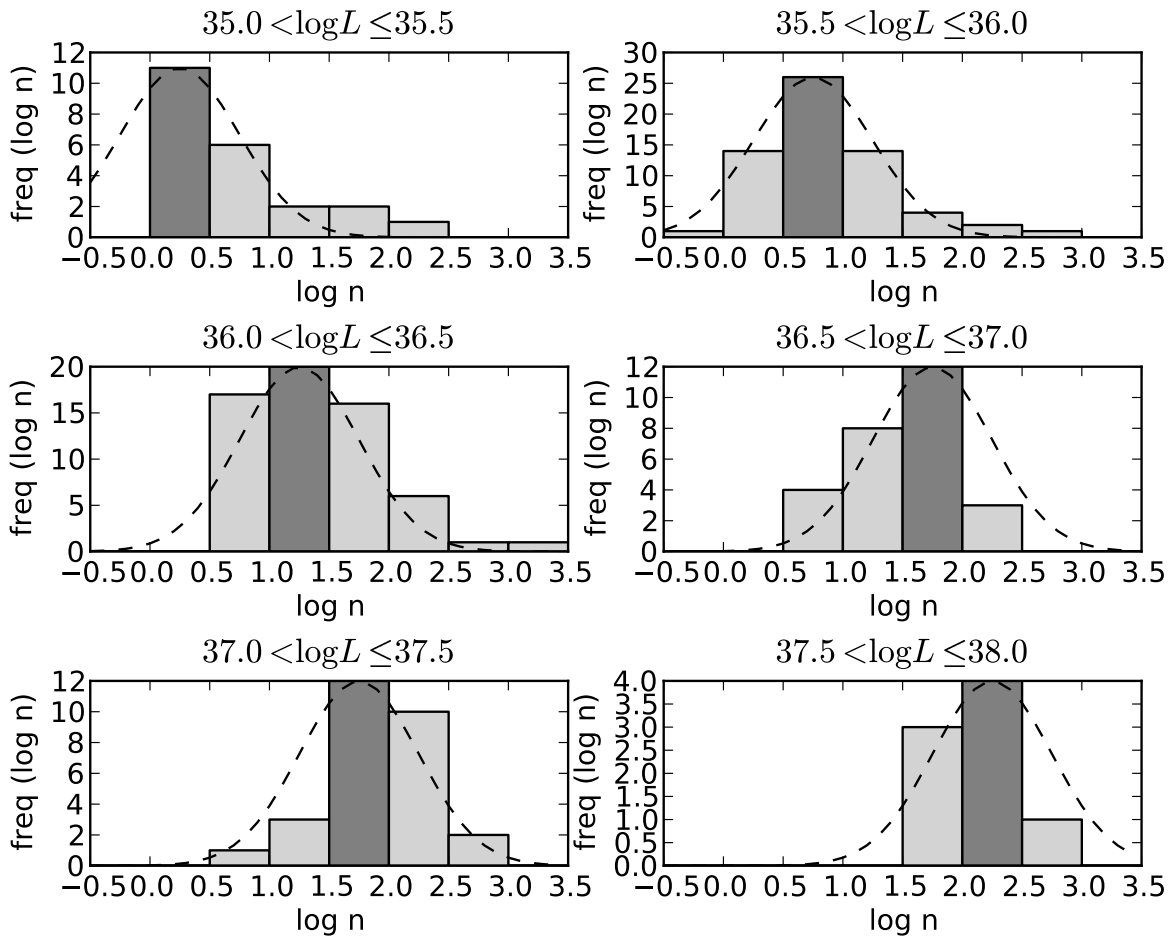}
  \caption{\label{fig:histbins} The frequency of occurrence of log($n$) in each range of the (log) far-UV luminosity is plotted here. The bin with the highest frequency is dark gray. Selection effects cut off the data points on the left side and skew the histograms towards the right. In order to illustrate a (log-)normal distribution which might be expected from stochastic noise, dashed lines represent Gaussian distributions of an arbitrary height and width centered at the highest frequency bars.}
\end{figure*}

The appearance of an optical depth limit to the 21-cm surface brightness leads to a serious bias in determining the true exponent of the star formation law from the observations. The dash-dot OLS fit line in the left panel of Figure \ref{fig:m33sfr} clearly shows a slope (of 0.6) which is too low, a consequence of biasing the calculated gas density to values that are artificially high. To explore this selection effect further, we divided the $L_{UV}$ data into six (log) luminosity ranges, and computed the histograms of the $\log n$ values in these six horizontal strips; the results are shown in Figure \ref{fig:histbins}. These distributions appear to be approximately log-normal, but are also clearly biased by the selection effect. As a first attempt to correct for this bias, we identified the value of $\log n$ corresponding to the peak of the histogram in each strip (shaded darker gray in the figure) and fitted an OLS line to these six values. The exponent found by this (oversimplified) method is 1.2 (graphed as the dashed line labeled ``Hist'' in the right panel of figure \ref{fig:m33sfr}). This value must still be too small, since we have assumed that the histograms of figure \ref{fig:histbins} are merely cut off to varying degrees on their left sides, rather than being cut off \textit{and skewed} towards their right sides in some nonlinear way by the bias. Removing the outer bins does not change the slope.

%Figure positioned manually

\subsection{A regression analysis}
\label{sec:regression}

We have shown that selection effects truncate and skew the observed distribution of cloud densities, biasing the value of the exponent obtained by a simple least-squares fit to the star formation law to values that are too low. These observational limits cause selection effects similar to the well-known ``Malmquist bias''. In order to obtain a more statistically-sound value for the exponent $\alpha$, a regression analysis based on the use of ``Monte Carlo Markov Chains'' (MCMC) can be used on the cloud density and the UV luminosity data (Equation \ref{eqn:n}), taking into account both the bias and the (relatively large) uncertainties in our cloud density measurements. This form of Bayesian inference \citep[see e.g.][]{1996ApJ...462..672T}, uses \textit{a priori} assumptions of the data, that are adjusted by the measured data. It provides a statistical formalism to obtain the true exponent based on simulations of unbiased data, where the ``simple approach'' in our previous section provided an intuitive way to estimate the missing data with similar assumptions. In both cases the assumption is the log-normal distribution of the values of $n$ in limited ranges of the UV luminosity.

The full procedure is described in \citet{2007ApJ...665.1489K}, where it is shown how this approach (referred to as a maximum likelihood estimate MLE using a Gaussian mixture model) is robust in the presence of censored data and large measurement errors. We used the IDL routines described in that paper to perform our analysis. The result of the MLE is an \textit{a posteriori} median estimate, which means that the slope (and intersect) resulting from this method are direct measurements of their true values.

The IDL routine constructs a Monte Carlo Markov Chain \citep[][review the formalism in their Appendices]{2009ApJ...697..258J}, where new data points are simulated using the ``Metropolis - Hastings'' algorithm. This particular algorithm is appropriate when errors dominate, and when the variables may not be independent, as is the case here. The regression parameters (generally) converge after several thousand iterations to provide the parameters drawn from the posterior distribution. It is important to realize that the value for the exponent in the star formation law obtained in this way is a random draw from the true distribution of all exponents consistent with the noisy and censored input data. The method can therefore also be used repeatedly to provide an estimate of the stochastic error in the exponent. 

\begin{figure}[ht!]
  \centering
  \includegraphics[width=\linewidth]{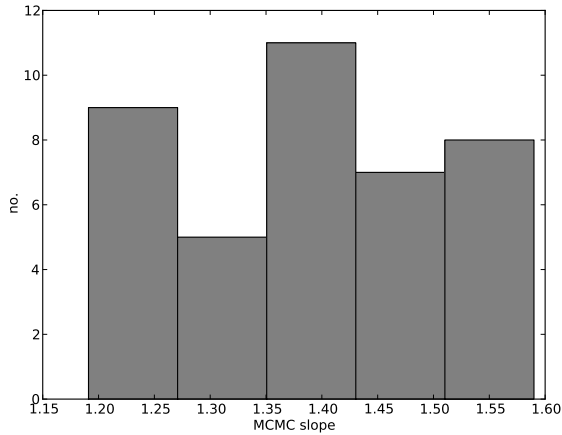}
  \caption{\label{fig:slopehist} The distribution of slopes obtained by applying the MCMC method 40 times is shown in this histogram. The most common value is 1.4, followed by 1.2, while a range of values between 1.2 and 1.6 is obtained.}
\end{figure}

To use the MCMC method as presented by \citet{2007ApJ...665.1489K}, we need to adopt reasonable estimates for the ($1 \sigma$) errors in $\log(n)$ and $\log(L_{UV})$. This is essential for the algorithm to estimate the true distribution of $n$ and $L_{UV}$. The crude but intuitive attempt at correcting for the selection effects described in the previous section shows that the cloud densities (ignoring the truncation and skewing) roughly follow a log-normal distribution. Also, the errors in $n$ dominate the correlation between log($n$) and log($L_{UV}$). We take absolute uncertainties in the values of log($n$) and log($L_{UV}$) of 0.5 and 0.2 respectively, conservative values reflecting the relatively large error in $n$ and the smaller error in $L$, where $L$ is directly based on the measured UV flux\footnote{Our forthcoming paper on the input data set will have the values of observed UV fluxes.}. We then run the regression routines repeatedly. The resulting slopes are shown in Figure \ref{fig:slopehist}. The distribution of the values of the slope is non-normal, but the average value is 1.4 with a range of $\pm 0.2$.
Note that the MCMC method is invariant to swapping the coordinate axes. It is also robust against small variations in the estimated uncertainties for $n$ and $L_{UV}$, and setting these uncertainties to nearly 0 brings back the OLS bias. The solid line with slope $1.4$ labeled MCMC in the right panel of Figure \ref{fig:m33sfr} reflects the outcome of our Bayesian approach to correct for our selection effects, and provides our best estimate of the exponent in the Schmidt Law of star formation.

%Figure positioned manually

\section{Discussion and conclusions}
\label{sec:conclusions}

The new observational method to estimate the slope of the star formation law that we present here can be further improved in various ways. While some selection effects may have been dealt with acceptably, others remain, and all such effects need to be considered carefully as we have attempted to do here. 

\begin{enumerate}
\item The method yields cloud densities with a relatively high uncertainty (we used a conservative estimate of 50\%). This uncertainty could be reduced by improving the quality of the data; however, the spread in cloud densities may very well be intrinsic, similar to the scatter in the local metallicities measured at different places in the disks of nearby galaxies \citep[e.g.\ ][]{2008ApJ...675.1213R}.
\item The star formation rate is derived from the UV fluxes of OB associations, and these may be affected by extinction to various degrees which we have ignored here. A detailed extinction map of M33 would be helpful, as only a spatially-varying extinction correction would influence our determination of the exponent in the star formation law.
\item A better calibration of the observed UV flux to the actual star formation rate could be carried out. However, \citet{1998ApJ...498..106M} found that the star formation rate is directly proportional to the FUV luminosity at 150 nm, which is very close to the effective wavelength of the \textit{GALEX} far-UV data, so it is unlikely that further precision here would change our results.
\item A correction for the effects of high optical depth in the 21-cm line could be developed in order to obtain more accurate estimates of the \mbox{H{\sc i}} column density. While such a correction would surely help, it is unfortunately not clear how one might determine it.
\item Lower atomic hydrogen column densities can only be used if they can be distinguished clearly from the general \mbox{H{\sc i}} background. More sensitivity and high spatial resolution would be required here, although we have argued that the biasing effect of limited \mbox{H{\sc i}} sensitivity on the power law slope is minor.
\end{enumerate}

Unique to our approach is the use of cloud volume densities, not surface densities, and the enhanced sensitivity of the PDR method to lower gas densities (less than a few hundred baryons $\rm{cm}^{-3}$) when compared to the higher-density sensitivity of the CO(1-0) estimates. To our knowledge, the only exception to this is the work of \citet{2008ARep...52..257A}; these authors specifically aimed to explore a volume density Schmidt Law. However, they could make only rough (CO-based) gas volume density estimates, based on column density measurements and an (uncertain) theoretical model of the thickness of galactic disks.

To conclude, we have carried out the first direct determination of the exponent in the star formation law as initially formulated by Schmidt over 50 years ago. The star formation rate has been estimated using the far-UV luminosities of OB associations in M33. The total gas volume density (atomic + molecular) has been obtained using a method which regards the atomic hydrogen as the dissociated ``skins'' of the molecular clouds out of which the young stars have formed. We have used a simple one-dimensional slab model for the dissociated layer of \mbox{H{\sc i}} on the cloud surfaces. We also show how observational selection leads to a serious bias in the determination of the exponent of the star formation law. We have used a ``maximum likelihood'' technique to account for this selection bias. Our result for the slope of the star formation law in M33 is $1.4 \pm 0.2$. We note that in this paper we  have addressed only the question of the value of the exponent in the Schmidt Law, and not the normalization. Application of our method to a larger sample of galaxies will provide additional information on this normalization and its universality.

Finally we point out that the value we have obtained for the exponent in the Schmidt Law for star formation is close to that used in many recent successful numerical simulations of galaxy formation and evolution. This correspondence lends further credence to our view that the \mbox{H{\sc i}} in galaxy disks is a \textit{product} of the star formation process, and that this process can be usefully analysed using the simple physical model described in this paper. 

\begin{acknowledgements}
JSH is grateful for a postdoctoral fellowship from the \textit{Centre de Recherche en Astrophysique du Qu\'{e}bec} (CRAQ) at Laval University, where the data were analysed. RJA acknowledges helpful suggestions on the text of this paper from Frank Shu, Jim Pringle, Colin Norman, and David Neufeld. We also thank the anonymous referee for suggestions that improved this paper.
\end{acknowledgements}

\end{document}